\documentclass[%
 reprint,
 amsmath,amssymb,bm
 aps,
]{revtex4-2}

\usepackage{algorithm}
\usepackage{algpseudocode}
\usepackage{graphicx}
\usepackage{dcolumn}
\usepackage{bm,color}
\usepackage{scalerel}

\begin{document}

\preprint{APS/123-QED}

\title{Capturing membrane structure and function in lattice Boltzmann models}
\thanks{A footnote to the article title}%

\author{James E. McClure}
\affiliation{%
National Security Institute and Center for Soft Matter and
Biological Physics \\
Virginia Polytechnic and State University \\
Blacksburg, VA, 24060, USA
}
 \email{mcclurej@vt.edu}

\author{Zhe Li}
\affiliation{%
Research School of Physics\\
The Australian National University\\
Canberra, 2601, Australia
}

\date{\today}

\begin{abstract}
We develop a mesoscopic approach to model the non-equilibrium behavior of membranes at the cellular scale.
Relying on lattice Boltzmann methods, we develop a solution procedure to recover the Nernst-Planck equations and Gauss's law. A general closure rule is developed to describe mass transport across the membrane, which is able to account for protein-mediated diffusion based on a coarse-grained representation.
We demonstrate that our model is able to recover the Goldman equation  from first principles and show that hyper-polarization occurs when membrane charging dynamics are controlled by multiple relaxation timescales. The approach provides a promising way to characterize non-equilibrium behaviors that arise due to the role of membranes in mediating transport based on realistic three-dimensional cell geometries. 
\end{abstract}

\maketitle

Membranes play an essential role in biological systems, providing the basic conditions that allow cells to control the movement of material into and out of the cell. 
Ion transport is particularly important within this context, since this is used to control electrical responses and sense changes in the extra-cellular environment \citep{doi:10.1063/5.0033608}. Key cellular functions depend on how the actions of
membrane proteins interact with the overall cell structure \citep{Teeter2018,Dutta2017,Yang2015}. More generally, cell geometry will always play a key role in electrical signaling due to well-known effects of confinement on diffusion \citep{BARHOUM20161,MIKA2011117,doi:10.1091/mbc.E15-04-0186}. 
Recently, experimental imaging techniques to resolve the geometric structure of real cells have become more widespread \citep{10.2307/24101088,Thorn_2016,doi:10.1126/science.1251652,Elliott_2020}. Within this context, mesoscopic simulation methods provide an intriguing way to quantitatively explore non-equilibrium behaviors based on true-to-life geometric constraints and biologically relevant time scales. Mesoscopic methods are constructed based on a coarse-grained representation of the molecular physics, allowing for significantly larger timesteps and spatial domain sizes as compared with molecular dynamics techniques \citep{10.1145/3458817.3487397}. At the same time, mesoscopic methods are able to capture more detailed information about non-equilibrium processes, well beyond what is possible with simplified rule based-models \citep{FITZHUGH1961445}. These considerations suggest that mesoscopic simulation can fill an important gap with respect to the modeling of biological systems.

Direct observations of cell structure provide key motivation for whole-cell models. Sources of 3D image data for biological cells are widely available
\citep{Thorn_2016,Fang2021,doi:10.1063/5.0082799,doi:10.1126/sciadv.1602580,doi:10.1063/5.0082799}. Many factors influence 
image data quality, and experimental trade-offs with respect to 
signal-to-noise ratio, resolution, and speed must be balanced to 
obtain good data \citep{LEMON202034}. The input for simulation workflows
will generally be the output for experimental workflows, consisting of
segmented structures that clearly identify the cell location and material constituents. Optical microscopy techniques present a wide range of opportunities to directly observe cellular geometry as well as resolve dynamic responses due to ion transport \citep{CLOUGH201995,Ferri_etal_2018}. 
The timescale to acquire image data using optical techniques can vary from milliseconds to minutes, depending on the particular technique and experimental objectives \citep{Fang2021,MANGEAT2021100009,Godin_etal_2014}. 
These experimental scales constrain the types of processes that can be observed,
since the timescale for biological phenomena can vary widely.
For neurons, the timescale to observe an action potential requires $\sim 1$ ms resolution and the timescale for calcium influx requires $\sim 100$ ms.
The length and time scales required to model these phenomena are a natural match with mesoscopic methods. Furthermore, experimental imaging techniques for live cells are hindered by the destructive influence of photons; excessive thermal stimulation can undermine the integrity of cellular structure and alter dynamic behaviors \citep{doi:10.1126/science.1082160,Jensen_2013,10.1242/jcs.033837}. 
Physics-based simulation thereby provides a natural complement to experimental imaging, since it is a mechanism to non-destructively infer non-equilibrium behaviors at the appropriate timescale, resolving critical physics that are difficult or impossible to access experimentally. Opportunities to combine simulation with experimental imaging protocols represent an intriguing avenue to improve understanding of the relationship between biological structure and function. 

We develop a model to account for two main factors that
control the behavior of biological membranes. First is the membrane geometry, which is determined by the cell structure. Second are the diffusion characteristics for the membrane, which are determined by the membrane composition and interactions between membrane proteins and the local environment. In principle, the cell geometry and membrane transport properties can be independently tuned to design membranes with desired properties and optimize
their performance. Within this context, mesoscopic
simulation provides a natural way to perform studies that would be 
prohibitively expensive and time-consuming in a laboratory setting. 
Our approach focuses on the role of membranes as barriers to diffusion, noting that ion transport across the lipid bilayer is of particular interest due to the central role of electrochemical gradients in many basic biological functions. We propose straightforward mesoscopic closure rules to couple the representation of the membrane to the underlying molecular transport mechanisms based on experimentally observable phenomena.  Our approach is constructed as a complementary tool for experimental protocols, such that images from different sources can be easily ingested into the simulator. Noting that accurate representation for the spatial distribution of charge is important to correctly model the non-equilibrium behavior for the membrane, the proposed solution procedure offers a first-principles approach to model the effects of ion transport at the whole-cell level. We verify that the described approach is able to model both equilibrium and non-equilibrium membrane responses, and show that it can recover both the Nernst reversal potential and the Goldman potential.

\begin{figure}[b]
\includegraphics[width=1.0\linewidth]{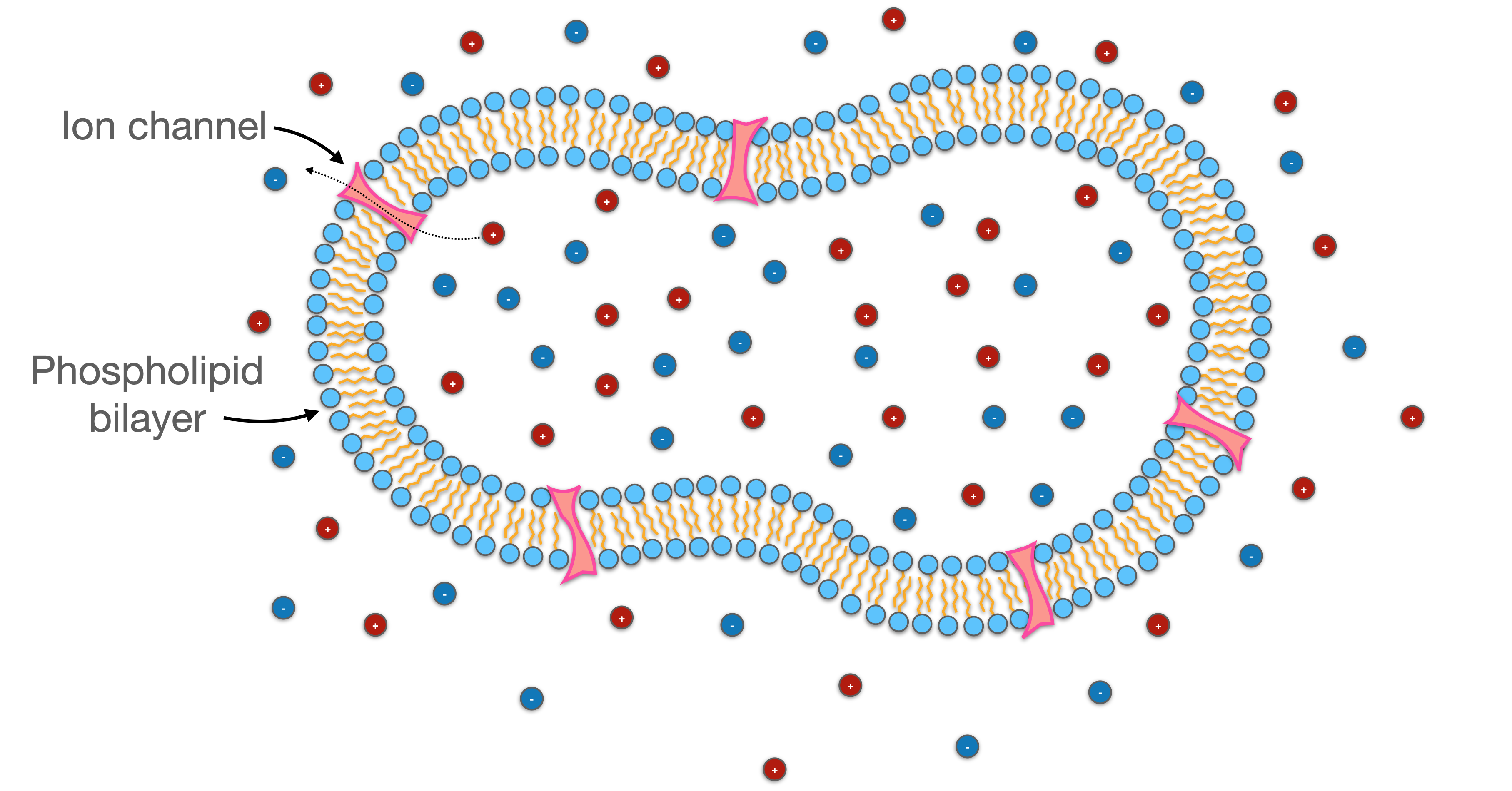}
\caption{\label{fig:cell} Structure of the phospholipid bilayer membrane and role in ion transport within a cell. The membrane serves as a barrier that prevents ions from moving
freely between the cell and the extracellular fluid. Transport across the membrane is
controlled by ion channels by protein-mediated diffusion.
}
\end{figure}

\section{Methods}

Mechanisms that control mass transport across membranes are responsible for many interesting non-equilibrium dynamics. Proteins embedded in biological membranes contribute to mass transport in novel ways, pumping ions against the electrochemical gradient, and relying on various signaling pathways to trigger voltage-gated ion channels that rapidly discharge the membrane capacitance \citep{doi:10.1146/annurev.biochem.77.062706.154450}. At the mesoscopic scale these effects can be modeled by controlling the direction-dependent membrane permeability, noting that the membrane properties may vary in space (e.g. due to the protein composition within the membrane) as well as time (e.g. due to dynamically triggered effects). Mass transport across the membrane is coupled to diffusion in the cell exterior and interior, and a complete model must include both aspects of the physics. Here we develop a generic mesoscopic formulation to model these effects, relying on three fundamental pieces:
(1) a model for electro-diffusion in the bulk regions;
(2) data structures to represent the membrane geometry; and
(3) a closure rule to predict mass transport across the membrane according to customizable rules.

\subsection{Lattice Boltzmann Formulation}

Lattice Boltzmann methods (LBMs) are a flexible class of numerical method that are capable of modeling a wide range of complex transport phenomena \citep{PhysRevE.56.6811,shan_yuan_chen_2006}. 
Due to favorable computational characteristics, LBMs are able to model mesoscopic dynamics at length and timescales that are presently inaccessible to any other numerical method. 
In terms of the scale of representation, LBMs are most similar to finite-element methods (FEM) and finite-volume methods (FVM) \citep{10.3389/fninf.2020.00011,doi:10.1063/5.0033608,Mori_Peskin_2009}. 
LBMs differ from these approaches because closure rules are typically developed from coarse-grained quasi-molecular interaction models rather than from continuum approximations. 
Furthermore, because the LBM is a discrete solution to the Boltzmann transport equation, 
non-ergodic transport behaviours can be treated 
based on quasi-molecular rules developed from a lower level in the modeling hierarchy, relying on fewer assumptions about the underlying processes.  

A model to recover the non-equilibrium electrical behavior within a cell
must describe the transport of charged chemical species as well
as account for their effect on the electric field and associated forces.
The electric potential $\psi$ satisfies the differential form of Gauss's law,
\begin{equation}
    \nabla^2 \psi = -\frac{\rho_e}{\epsilon_r \epsilon_0} \;,
    \label{eq:Poisson-Boltzmann}
\end{equation}
where $\epsilon_0$ is the permittivity of free space and
$\epsilon_r$ is the material-dependent relative permittivity.
Solution of Eq. \ref{eq:Poisson-Boltzmann} is coupled to ion transport
based on  the charge density $\rho_e$, which can be computed directly  based on the distribution of ions in the system,
\begin{equation}
    \rho_e = \sum_k F z_k C_k\;,
    \label{eq:charge-density}
\end{equation}
where Faraday's constant is given by $F=eN_A = 96485$ C/mol, $z_k$ is the valence charge of $k$th ion species, $C_k$ is the associated concentration, and $N_A$ is the Avogadro constant. 
The evolution for each ion concentration $C_k$ satisfies the Nernst-Planck equations,
\begin{eqnarray}
    && \frac{\partial C_k}{\partial t} + \nabla \cdot \mathbf{j}_k = 0\;,  \\ 
    && \mathbf{j}_k = C_k \mathbf{u} - D_k \Big( \nabla C_k + \frac{z_k C_k}{V_T} \nabla \psi\Big)\;, 
    \label{eq:Nernst-Planck}
\end{eqnarray}
where $D_k$ is the diffusion coefficient for ion $k$. 
The thermal voltage is defined as $V_T=RT/F$ based on the ideal gas constant $R$ and temperature $T$.
The mass flux $\mathbf{j}_k$ includes contributions from advection
as well as from the gradients in the chemical and electric potential. 
The velocity $\mathbf{u}$ can be determined from separate solution of
a momentum equation. For the cases considered in this work
$\mathbf{u}$ is set to zero for simplicity, on the basis that mass of  ions is small compared to the mass of the electrolyte. 

The solution procedure for Eqs. \ref{eq:Poisson-Boltzmann} and
\ref{eq:Nernst-Planck} is defined based on the lattice Boltzmann method. The LBM is constructed as an approximation to the continuous Boltzmann transport equation, utilizing Gauss-Hermite quadrature to formulate a discrete representation for the molecular velocity distribution \citep{shan_yuan_chen_2006}. 
To develop a solution procedure for Eq. \ref{eq:Poisson-Boltzmann}, we rely on
a three dimensional approximation involving nineteen discrete velocities (D3Q19), given by
\begin{eqnarray}
\bm{\xi}_{q} &=& \left\{ 
\begin{array}{r} 0 \\ 0 \\ 0 \end{array}
\begin{array}{r} 1 \\ 0 \\ 0 \end{array}
\begin{array}{r} -1 \\ 0 \\ 0 \end{array}
\begin{array}{r} 0 \\ 1 \\ 0 \end{array}
\begin{array}{r} 0 \\ -1 \\ 0 \end{array}
\begin{array}{r} 0 \\ 0 \\ 1 \end{array}
\begin{array}{r} 0 \\ 0 \\ -1 \end{array} 
\begin{array}{r} 1 \\ 1 \\ 0 \end{array}
\begin{array}{r} -1 \\ -1 \\ 0 \end{array}
\begin{array}{r} 1 \\ -1 \\ 0 \end{array}
\begin{array}{r} -1 \\ 1 \\ 0 \end{array}
\right .
\nonumber \\
&&
\left .
\begin{array}{r} 1 \\ 0 \\ 1  \end{array}
\begin{array}{r} -1 \\ 0 \\ -1  \end{array}
\begin{array}{r} 1 \\ 0\\ -1  \end{array}
\begin{array}{r} -1 \\ 0\\ 1  \end{array}
\begin{array}{r} 0\\ 1 \\ 1  \end{array}
\begin{array}{r} 0\\ -1 \\ -1  \end{array}
\begin{array}{r} 0\\ 1 \\ -1  \end{array}
\begin{array}{r} 0\\ -1 \\ 1  \end{array}
\right \} \; \frac{\Delta x}{\Delta t}. \nonumber
\label{eq:discrete-velocity}
\end{eqnarray}
the quadrature scheme leads to an associated set of discrete distributions
$g_q$, $q=0,1,\ldots,19$. The discrete representation can be used to construct efficient models for a wide range physics, including implementations that provide numerical solution of Poisson's equation \citep{WANG2006729,WANG2008575,WANG2010728,CHEN2014186,ZHANG2017219}. In our case,
effects of anisotropy can be significant due to the shape of the membrane;
a particular choice of weights is selected to minimize these effects.
The weights associated with a D3Q19 approximation to the Laplacian
are given by \cite{OReilly2006AFO},
\begin{eqnarray}
    \nabla^2_{fe} \psi (\mathbf{x}_i) &=& \frac{1}{6 \Delta x^2}
    \Bigg( 2 \sum_{q=1}^{6} \psi(\mathbf{x}_i + \bm{\xi}_q \Delta t) \nonumber \\ 
     && +  \sum_{q=7}^{18} \psi(\mathbf{x}_i + \bm{\xi}_q \Delta t)
   - 24 \psi (\mathbf{x}_i) \Bigg) \;.
\end{eqnarray}
Consistent with this approximation, we define the equilibrium functions
\begin{equation}
    g_q^{eq} =  w_q \psi\;, \quad   w_q =  \left \{ 
    \begin{array}{cl}
    \frac {1}{2} & \mbox{for $q=0$} \\
    \frac {1}{24} & \mbox{for $q=1,\ldots,6$}  \\
    \frac {1}{48} & \mbox{for $q=7,\ldots,18$} \\
    \end{array}
    \right.
    \label{eq:geq}
\end{equation}
which implies that 
\begin{equation}
    \psi = \sum_{q=0}^{Q} g_q^{eq}\;.
\end{equation}
Given a particular initial condition for $\psi$, let us consider application of the standard D3Q19 streaming step based on the equilibrium distributions
\begin{equation}
    g_q^\prime(\mathbf{x}, t) = g_q^{eq}(\mathbf{x}-\bm{\xi}_q\Delta t, t+ \Delta t)\;.
\end{equation}
Due to the choice of weights, after streaming an approximation to the Laplacian
is easily obtained
\begin{equation}
  \nabla^2_{fe} \psi (\mathbf{x}_i) = 
8 \Big[ -g_0 +  \sum_{q=1}^Q g_q^\prime(\mathbf{x}, t) \Big] \;,
\end{equation}
Relative to the solution of Gauss's law, the error is given by
\begin{equation}
  \varepsilon_{\psi} = 
8 \Big[ -g_0 +  \sum_{q=1}^Q g_q^\prime(\mathbf{x}, t) \Big] 
+ \frac{\rho_e}{\epsilon_r \epsilon_0} \;.
\end{equation}
Using the fact that $f_0 = W_0 \psi$, we can compute the value 
$\psi^\prime$ that would kill the error. We set $\varepsilon_{\psi}=0$
and rearrange terms to obtain
\begin{equation}
\psi^\prime (\mathbf{x},t) = \frac{1}{W_0}\Big[   \sum_{q=1}^Q g_q^\prime(\mathbf{x}, t) 
+ \frac{1}{8}\frac{\rho_e}{\epsilon_r \epsilon_0}\Big]  \;.
\end{equation}
The local value of the potential is then updated based on a relaxation scheme, which is controlled by the relaxation time $\tau_\psi$
\begin{equation}
\psi(\mathbf{x},t+\Delta t) \leftarrow \Big(1 - \frac{1}{\tau_\psi} \Big )\psi (\mathbf{x},t)
+ \frac{1}{\tau_\psi} \psi^\prime (\mathbf{x},t)\;.
\end{equation}
The algorithm can then proceed to the next timestep based on 
Eq. \ref{eq:geq}.

A LBM solution to recover Eq. \ref{eq:Nernst-Planck} is developed using a three-dimensional, seven velocity (D3Q7) lattice structure, which correspond to $q=0,1,\ldots,6$ from
Eq. \ref{eq:discrete-velocity}. Each distribution is associated with a particular discrete velocity, $f^k_q$. The concentration is given  by their sum,
\begin{equation}
    C_k  = \sum_{q=0}^{6} f^k_q \;.
\end{equation}
Lattice Boltzmann equations (LBEs) are defined to determine the
evolution of the distributions $f_q^k$, 
\begin{equation}
    f^{k}_q (\mathbf{x}_n + \bm{\xi}_q \Delta t, t+ \Delta t)-
        f^{k}_q (\mathbf{x}_n, t) = \frac{1}{\lambda_k} 
        \Big( f^{k}_q - f^{eq}_q \Big)\;,
             \label{eq:mass-LBE}
\end{equation}
where the relaxation time $\lambda_k$ controls the bulk diffusion coefficient,
\begin{equation}
    D_k = c_s^2\Big( \lambda_k - \frac 12\Big)\;.
\end{equation}
The speed of sound for the D3Q7 lattice model is $c_s^2 = \frac 14$ and the weights are $W_0 = 1/4$ and $W_1,\ldots, W_6 = 1/8$.
Equilibrium distributions are established from the fact that molecular velocity distribution follows a Gaussian distribution within the bulk fluids,
\begin{equation}
          f^{eq}_q = W_q C_k \Big[ 1 + \frac{\bm{\xi_q}\cdot \mathbf{u}^\prime}{c_s^2} \Big]\;.
\label{eq:LBE-equilibrium}
\end{equation}
The velocity $\mathbf{u}^\prime$ is given by
\begin{equation}
    \mathbf{u}^\prime = \mathbf{u} - \frac{z_k D_k}{V_T} \nabla \psi \;.
    \label{eq:LBE-u}
\end{equation}
Solution of Eqs. \ref{eq:mass-LBE}--\ref{eq:LBE-u} will recover the Nernst-Planck Equations \citep{chapman1990mathematical}.
Combined with a numerical scheme to solve Eq. \ref{eq:Poisson-Boltzmann} this is sufficient to define a model for electrodiffusion within the bulk fluids
\citep{WANG2006729,WANG2008575,WANG2010728,CHEN2014186,ZHANG2017219}. Novel rules must be developed to model transport across the membrane, since
the local diffusion properties will differ substantially from the behavior within the  bulk fluids.


\subsection{Membrane representation and closure rule}

\begin{figure}[b]
\includegraphics[width=1.0\linewidth]{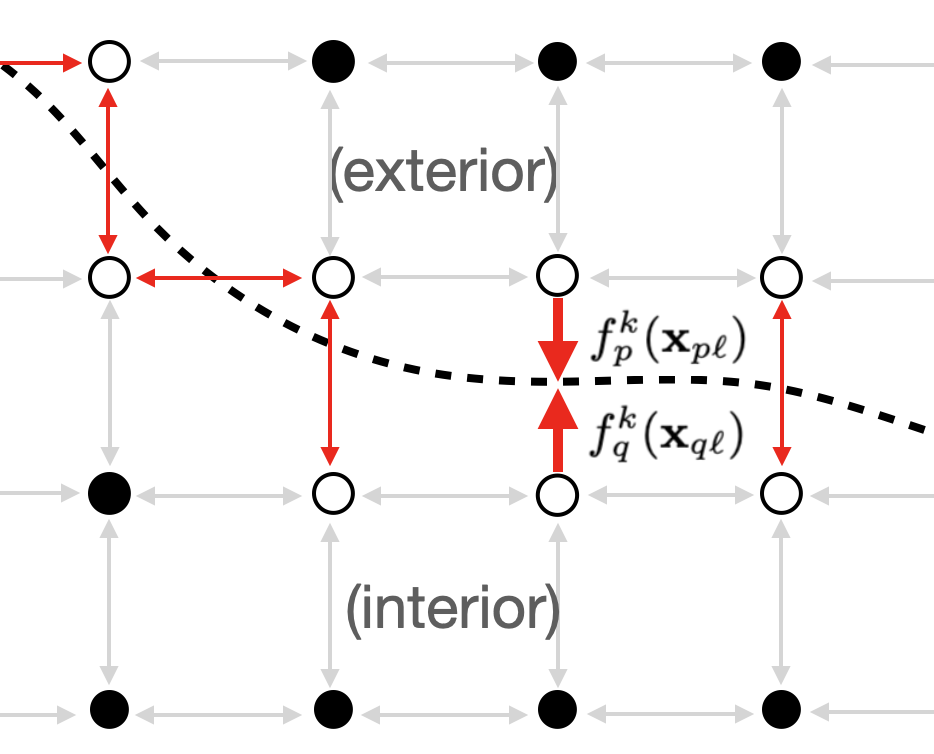}
\caption{\label{fig:membrane} Representation of a membrane (dashed-line) within lattice Boltzmann models: all discrete velocity links that cross the membrane are identified  and stored in a dedicated data structure. Transport behavior for membrane links (open sites and associated bold arrows)
are determined independently from the rules used for normal links (solid sites and light arrows). Distribution 
$f^k_q (\mathbf{x}_{q\ell})$ is on the interior side of the membrane, and $f^k_p (\mathbf{x}_{p\ell})$ is on the exterior side.
}
\end{figure}

The numerical representation for membrane structure is the critical factor to implement general whole-cell modeling capabilities. Due to scale separation it is advantageous to represent a biological membrane as a two-dimensional entity. The length scale for most cells falls within the range of $1$ $\mu$m to $1$ mm. In contrast, the thickness of the lipid bilayer is $~5$ nm, below the resolution for many relevant optical techniques. Accurately representing the membrane location can be accomplished by using signed distance functions, which are widely used in front-tracking algorithms. Given an observed structure, a distance map $\mathcal{D}(\mathbf{x})$ can be generated using approaches developed for level set methods \citep{Sethian1591}. The membrane location corresponds to $\mathcal{D}(\mathbf{x}) = 0$, which can be used to inform the lattice Boltzmann model. This is sufficient to capture the core function of the membrane as being a barrier to mixing within the system. Based on this representation, we may take advantage of the LBM data structures to enforce local rules that govern mass transport across the membrane. This is accomplished by considering the symmetry of the lattice based on the discrete velocity vectors $\bm{\xi}_q$, noting that for each distribution $f_q(\mathbf{x}_{q\ell},t)$ carrying mass out of the cell, there is an opposing distribution $f_{p}(\mathbf{x}_p,t)$ that will transport mass into to the cell, with $\mathbf{x}_p = \mathbf{x}_q+\bm{\xi}_q\Delta t$. This is depicted in Figure \ref{fig:membrane}. The standard lattice Boltzmann streaming step would propagate $f_q(\mathbf{x}_{q\ell},t)$  to site $\mathbf{x}_{p\ell}$ at time $t+\Delta t$, while $f_{p}(\mathbf{x}_{p\ell}, t)$ will propagate to site $\mathbf{x}_{q\ell}$. Based on the structure of the discrete velocity set, each distribution will be paired with exactly one other distribution, which we call a link. This symmetry can be exploited when considering the membrane representation. Links that cross the membrane can be determined by using the distance map to identify all lattice links that cross the membrane, as shown in Figure \ref{fig:membrane}. With the discrete velocity $\mathbf{\xi}_q$ pointing out of the membrane, membrane links $\ell$ uniquely satisfy the condition $\mathcal{D}(\mathbf{x}_{q\ell}) \mathcal{D}(\mathbf{x}_{p\ell}) < 0 $. The bounce-back rule is applied to exclude membrane links from participation in the streaming step \citep{Maier_1996}. A separate list is formed to store the membrane links, which are then treated according to an alternative closure rule. 
All mass transfer across the membrane occurs across these links. Manipulation of the lattice Boltzmann data structures is thereby able to incorporate the membrane geometry into the mesoscale representation.
Mass transport across these links can then be governed by local closure rules. 

\begin{algorithm}[H]
\caption{\label{alg:link-update} Membrane link update rule to govern mass transport of component $k$.}
\begin{algorithmic}
\For{$\ell=0,1,\ldots,L$}
    \State { $f_{q}^{k \prime} (\mathbf{x}_{q\ell})  \gets (1-\alpha^k_{\ell q}) f_{q}^{k} (\mathbf{x}_{q\ell}) + \alpha^k_{\ell p } f_{ p}^{k}  (\mathbf{x}_{p\ell})$ }
    \State {$f_{p}^{k \prime} (\mathbf{x}_{p\ell})  \gets (1-\alpha^k_{\ell p}) f_{p}^{k} (\mathbf{x}_{p\ell}) + \alpha^k_{\ell q } f_{q}^{k}  (\mathbf{x}_{q\ell})$}
\EndFor
\end{algorithmic}
\end{algorithm}

In this work we consider the membrane location to be static, since our primary objective is model mass diffusion across the membrane. The membrane inhibits mass transport and normal diffusion statistics may consequently break down in its vicinity. Behavior may be non-ergodic due to the role played by the membrane in preventing mixing in the system, and due to the anomalous diffusion properties associated with protein-mediated transport \citep{Weigel_etal_2011}.
It is therefore appropriate to formulate the link update rule in a way that allows mesoscopic rules to deviate from those dictated by equipartition of energy. A coarse-grained strategy is defined by considering the net rate of mass transport across the membrane over the simulation timestep $\Delta t$.
This can be embedded in the membrane permeability for each ion, which may vary with both
time and space. Since the permeability may also be direction-dependent, independent coefficients $0 \le \alpha^k_{\ell q } \le 1$ and $0\le \alpha^k_{\ell p}\le 1$ are defined to control the fraction of mass that crosses the membrane from either side. The mass flux out of the cell across link $\ell$ is
\begin{equation}
    \mathbf{j}_k^\ell  = \big(\alpha^k_{\ell q } f^k_q  - \alpha^k_{\ell p } f^k_p \big) \bm{\xi}_q \;.
    \label{eq:membrane-permeability}
\end{equation}
Algorithm 1 summarizes how the closure rule is implemented to move each ion
across the membrane. Since Eq. \ref{eq:membrane-permeability}
is tied to the physical transport of ions
across the surface, it has a straightforward interpretation that may
be easily generalized to describe a wide range of non-equilbrium membrane
transport behaviors. Rules can be customized to consider the direction-dependent
membrane permeability to vary as a function of the local voltage, concentration, etc.
\begin{equation}
     \alpha^k_{\ell q} =  \alpha^k_{\ell q} (\psi, C_j, \ldots)\;, \quad
     \alpha^k_{\ell p} =  \alpha^k_{\ell p} (\psi, C_j, \ldots)\;.
\end{equation}
To understand the relationship between the  coefficients and the local diffusion coefficient, 
consider the distribution $f_{q}^k(\mathbf{x}_{q\ell})$. Particles transported across the membrane
are displaced from $\mathbf{x}_{q\ell}$ by $\Delta \mathbf{x}_q = \bm{\xi}_q \Delta t$. The displacement for remaining particles is zero.
Since $\alpha^k_{\ell q}$ is the fraction of particles that cross
the membrane, the mean squared distance traveled
by particles associated with the direction of travel $\bm{\xi}_q$ is
\begin{equation}
  \mbox{MSD}_{\ell q} = \alpha^k_{\ell q} ( \Delta \mathbf{x}_q)^2 \;.
  \label{eq:MSD}
\end{equation}
The diffusion properties can therefore be controlled independently for
each link, and are de-coupled from the bulk diffusion coefficient $D_k$. The formulation can even be applied to consider the case where $\alpha^k_{\ell q}$ is different from $\alpha^k_{\ell p}$, which corresponds to the situation where ion $k$ is subjected to active transport rather than passive transport. The representation
is coarse-grained based on the timestep $\Delta t$, which will generally
be relatively large as compared to the molecular timescale. Approaches
to measure direction-dependent diffusion statistics in the vicinity of
the membrane (e.g. single molecule tracking \cite{Weron_etal_2017}) can thereby be linked to
Eq. \ref{eq:MSD} in a straightforward way.

\section{Results}

We consider multi-ion transport based on constant membrane permeability as a means to demonstrate the basic features of the developed model, and to demonstrate that the non-equilibrium electrical response can be recovered from first-principles. Ion $k$ diffuses across the membrane based on the effective diffusion coefficient for the membrane $\tilde{D}_k$,
which together with the membrane thickness $h$ defines the membrane 
permeability $p_k$. The coefficients are chosen such that bi-directional transport is unbiased, and capture the tendency for the membrane
to inhibit diffusion for ion $k$,
\begin{equation}
  \alpha^k_{\ell p } = \alpha^k_{\ell q } = \frac{\tilde{D}_k}{D_k}
   = \frac{{p}_k {h}}{D_k}\;.
   \label{eq:membrane-diffusion}
\end{equation}
This also provides the basis to choose the coefficients in Algorithm 1 
for situations where $\tilde{D}_k$ can be measured experimentally.
From a fundamental perspective, it is differences in the permeability of the membrane to different ions that produce the resting membrane potential \citep{10.1085/jgp.27.1.37}. In this situation, a gradient in ionic strength will spontaneously produce a membrane voltage. Under
stationary conditions, the concentration difference for each ion is determined by the membrane permeabilities. On this basis, the Goldman equation predicts the associated membrane potential. The familiar form for the Goldman equation considers three ion species, $\mbox{Na}^{\tiny{+}}$, $\mbox{K}^{\tiny{+}}$ and $\mbox{Cl}^{\tiny{-}}$, corresponding to the three most prevalent ions in cells. The Goldman potential $\psi^*$ can be predicted according to the associated permeabilities,
\begin{equation}
    \frac{\psi^*}{V_T}= \ln \frac {
    {p}_{\scaleto{\mbox{K}}{4pt}}{C_{\scaleto{\mbox{K}}{4pt}}^{*\scaleto{\mbox{(out)}}{5pt}}}
    +{p}_{\scaleto{\mbox{Na}}{4pt}}{C_{\scaleto{\mbox{Na}}{4pt}}^{*\scaleto{\mbox{(out)}}{5pt}}}
    +{p}_{\scaleto{\mbox{Cl}}{4pt}}{C_{\scaleto{\mbox{Cl}}{4pt}}^{*\scaleto{\mbox{(in)}}{5pt}}}
    }
    {    {p}_{\scaleto{\mbox{K}}{4pt}}{C_{\scaleto{\mbox{K}}{4pt}}^{*\scaleto{\mbox{(in)}}{5pt}}}
    +{p}_{\scaleto{\mbox{Na}}{4pt}}{C_{\scaleto{\mbox{Na}}{4pt}}^{*\scaleto{\mbox{(in)}}{5pt}}}
    +{p}_{\scaleto{\mbox{Cl}}{4pt}}{C_{\scaleto{\mbox{Cl}}{4pt}}^{*\scaleto{\mbox{(out)}}{5pt}}}
    }  \;.
    \label{eq:goldman}
\end{equation}
When the membrane is permeable only to a single ion $k$ (setting $p_j=0$ for $j\neq k$), the Goldman equation reduces to the Nernst potential,
\begin{equation}
  z_k \frac{\psi^*}{V_T} = \ln \frac{C_k^{*\scaleto{\mbox{(out)}}{5pt}}}{C_k^{*\scaleto{\mbox{(in)}}{5pt}}} \;.
    \label{eq:Nernst-potential}
\end{equation}

\begin{figure*}
\centering
\includegraphics[width=1\textwidth]{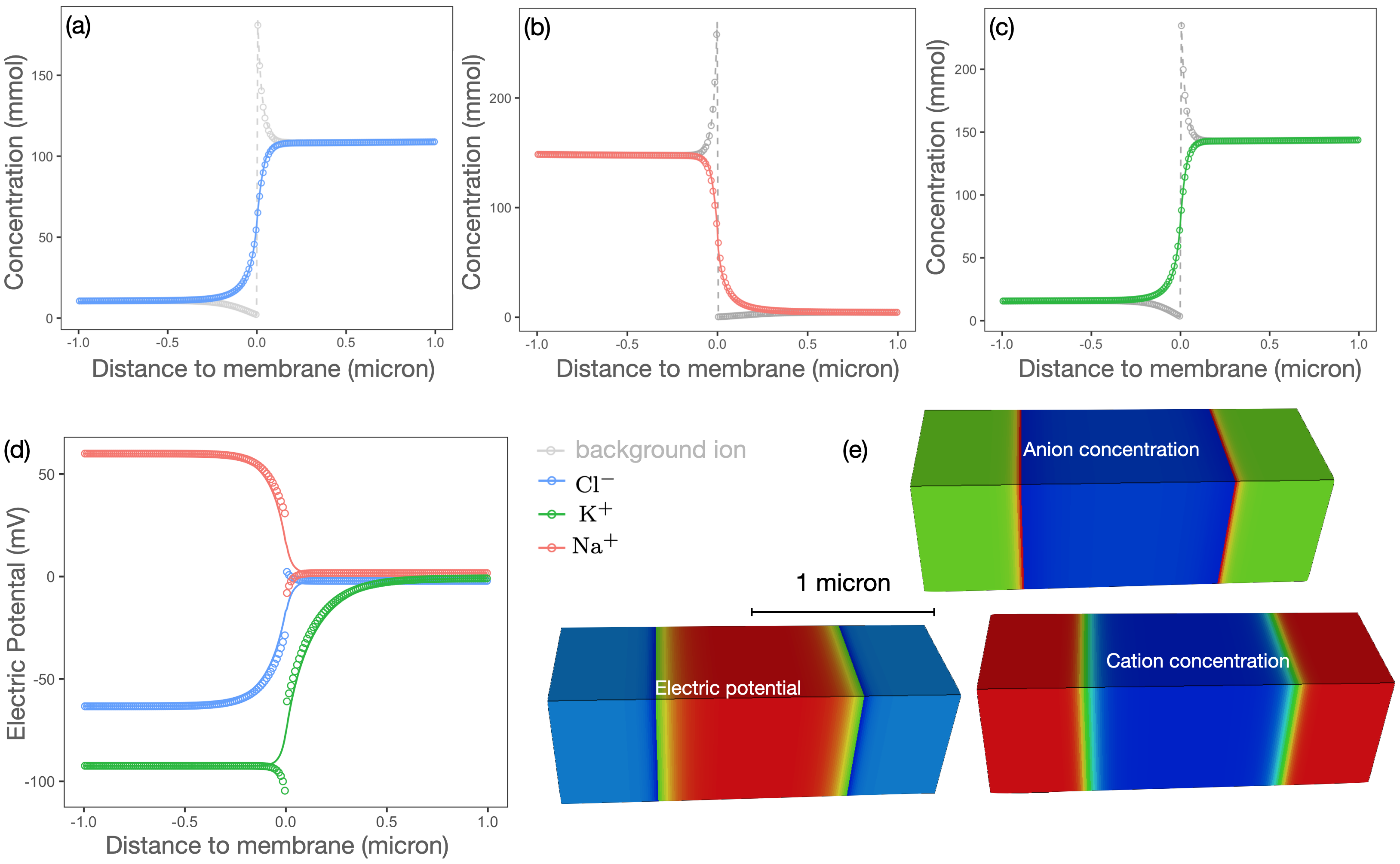}
\caption{(a)--(c) Ion concentration and (d) electric potential for a 1D membrane that is permeable to a single ion species. Simulation results are shown with the circle symbol, and analytical solution are solid lines. A background ion is used to initialize from an electroneutral condition on either side of membrane; (e) simulated electric potential and ion concentration fields. 
}\label{fig:Nernst}
\end{figure*}


To verify that expected stationary behavior is recovered, we simulate a planar membrane as shown in Figure \ref{fig:Nernst}. The initial ion concentration is piece-wise constant, with a different concentration on either side of the membrane. Three different simulations are considered, separately
considering the Nernst potential for $\mbox{Na}^{\tiny{+}}$, $\mbox{K}^{\tiny{+}}$ and $\mbox{Cl}^{\tiny{-}}$. Background ions with opposite charge are added to enforce electrical neutrality, meaning that there is a gradient in ionic strength across the membrane but no net charge in the simulation domain. The membrane is impermeable to the background ion, ${p}_{j} = 0$. With $p_k  > 0$ the ion $k$ is free to establish a stationary profile across the membrane. The permeable ion will diffuse across the membrane, driven by the gradient in chemical potential. Stationary profiles are obtained when diffusion is balanced by the drift current driven by the electric potential that is established as charges accumulate on both sides of membrane. Since the background ion is impermeable, it acts as a constraint on the permeable ion. Figure \ref{fig:Nernst}(a)--(d) shows the simulated ion concentration and electric potential plotted against the analytical solution reported by \cite{membrane_1d_analytical}. Ion concentration in the Gouy-Chapman layer closely matches the analytical solution. The ability to resolve the Guoy-Chapman layer is critical, since this region determines the surface charge density and plays a major role in determining the transient responses of the membrane. 

\begin{figure*}
\centering
\includegraphics[width=1\textwidth]{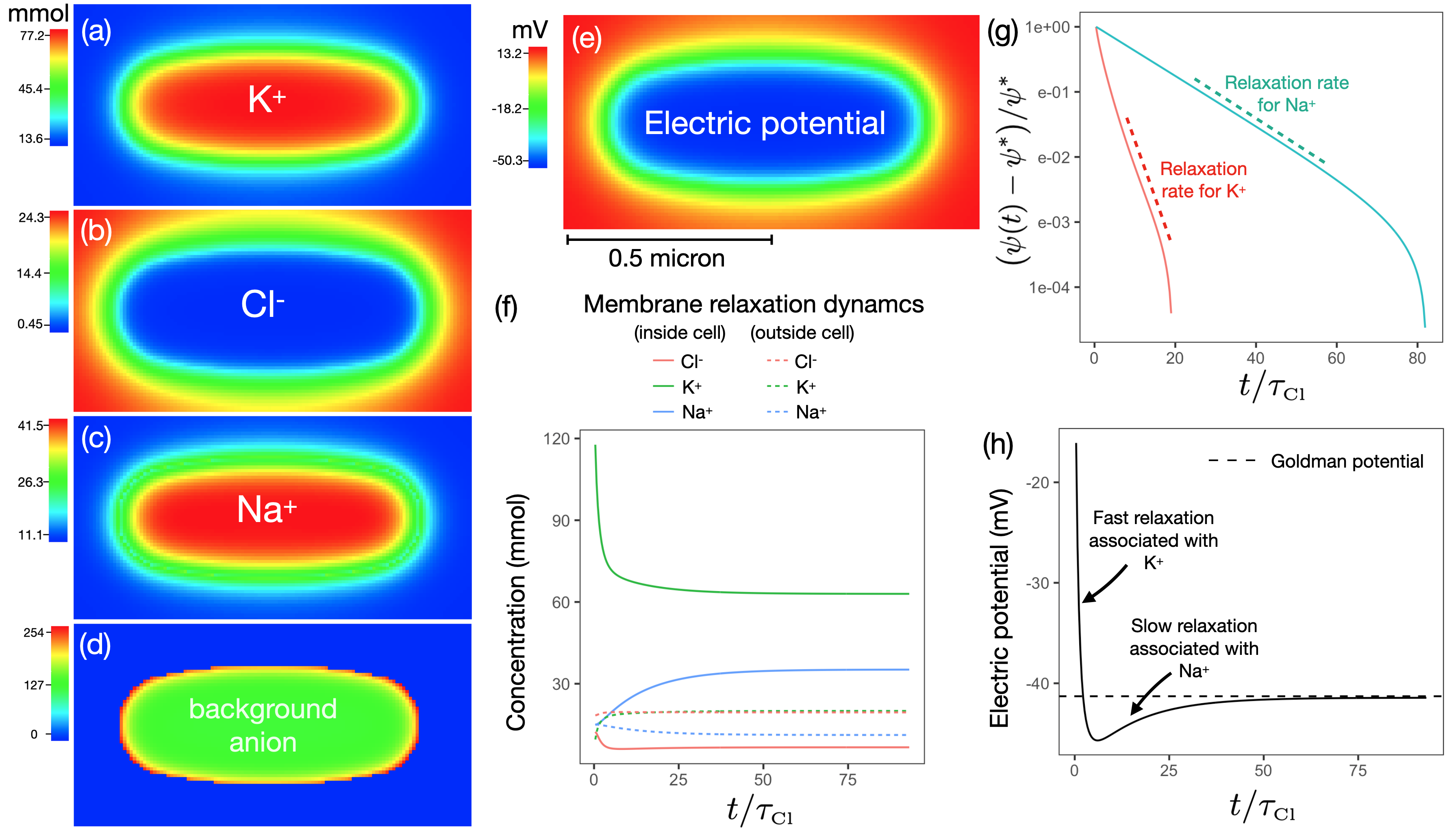}
\caption{Recovery of Goldman potential based on a simple membrane geometry:
(a)--(d) ion concentrations and 
(e) electric potential resulting from stationary conditions;
(f) concentration reaches stationary values following a multi-ion relaxation process;
(g) based on differing membrane permeability, the relaxation timescale for $\mbox{Na}^{\small{+}}$ is significantly slower than for $\mbox{K}^{\small{+}}$;
(h) the Goldman potential is recovered from the stationary solution,
with hyper-polarization due to the combined effect of the two
relaxation times.
}\label{fig:Goldman}
\end{figure*}

 Eqs. \ref{eq:goldman} and \ref{eq:Nernst-potential} are non-equilibrium expressions that describe a stationary state. They cannot be used to predict general non-equilibrium responses that occur based on ion transport. Mesoscopic simulation provides broad capabilities to model these dynamics. As an example, we simulate the charging dynamics due to diffusion of potassium, sodium and chloride ions based on a simple membrane shape, as shown in Figure \ref{fig:Goldman}(a)--(e). The length of the cell is $0.9$ $\mu$m and a maximum diameter of $0.36$ $\mu$m, which is comparable to the size for a typical bacteria. Simulations were performed on a 3D image of $128 \times 64 \times 64$ voxels with a resolution of $10$ nm. Initial concentrations were established based on the values listed in Table \ref{tab:table1}, with the membrane transport coefficients chosen based on Eq. \ref{eq:membrane-diffusion}. The diffusion coefficient $D_k = 1.0 \times 10^{-9}$ $\mbox{m}^2/\mbox{s}$ is applied for all ions $k$. Since the  relaxation process is dominated by the slower timescales associated with diffusion of $\mbox{Na}^{+}$ and $\mbox{K}^{+}$ across the membrane, the bulk diffusion coefficients play only a minor role in determining responses in the system. 

\begin{table}[h]
\begin{center}
\caption{Initial and final conditions for simulation results 
shown in Figure \ref{fig:Goldman}. The membrane permeability is directly proportional to $\tilde{D}_k$. Final concentration values 
are used to compute the Goldman equation. 
}\label{tab:table1} 
\begin{tabular}{@{}lccccc@{}}
\toprule
Ion &${\tilde{D}_k}/{D_k}$ 
&  $C_k^{\scaleto{\mbox{(in)}}{5pt}}(t_0)$  & $C_k^{*\scaleto{\mbox{(out)}}{5pt}}(t_0)$  &  $C_k^{*\scaleto{\mbox{(in)}}{5pt}}$  & $C_k^{*\scaleto{\mbox{(out)}}{5pt}}$ \\
\hline
$\mbox{Na}^{+}$  & $0.005$ & 15 mM & 20 mM  & 35.2 mM & 11.2 mM \\
$\mbox{K}^{+}$ & $0.1$ & 150 mM & 4 mM  & 63.0 mM & 2.0 mM \\
$\mbox{Cl}^{-}$  & $1.0$ & 10 mM & 16 mM & 6.7 mM & 19.5 mM \\
\botrule
\end{tabular}
\end{center}
\end{table}

For a membrane that is permeable to a single ion $k$, the charging dynamics can be predicted based on an associated relaxation time $\tau_k$,
\begin{equation}
%
\frac{\psi(t) - \psi^*}{\psi^*} =\exp{ \Big[\frac{t-t_0}{\tau_k}\Big]}\;,
\label{eq:membrane-relaxation}
\end{equation}
where the initial condition is given by $\psi(t_0) = 0$. The stationary cell
potential is $\psi^*$, with $t^*$ the time required to 
achieve stationary conditions, such that $\psi(t)=\psi^*$ for $t>t^*$. In the mesoscopic model, 
$\tau_k$ is controlled by the membrane permeability for ion $k$. When the membrane is permeable to multiple ions, multiple relaxation timescales will coexist. First principles simulation provides a natural mechanism to directly resolve these non-equilibrium physics, with minimal simplifying assumptions. 

A non-dimensional timescale is obtained by normalizing with respect to the
fastest relaxation rate, $\tau_{\scaleto{\mbox{Cl}}{4pt}}$. Taken independently, the
relaxation rate for the two positive ions is shown in Figure \ref{fig:Goldman}(f).
These rates are controlled by choosing the membrane permeability.
Given a particular membrane structure, the relaxation rate for ion $k$ can be determined by setting $p_j=0$ for $j \neq k$, retaining the same initial condition. The relaxation rates depicted in Figure \ref{fig:Goldman}(f) are obtained by combining this condition with the 
initial ion concentrations listed in Table \ref{tab:table1}.
For this situation Eq. \ref{eq:membrane-relaxation} will hold and the relaxation rate
is easily identified. For this system the relative relaxation rates are
$\tau_{\scaleto{\mbox{K}}{4pt}} / \tau_{\scaleto{\mbox{Cl}}{4pt}} = 2.2$ and $\tau_{\scaleto{\mbox{Na}}{4pt}} / \tau_{\scaleto{\mbox{Cl}}{4pt}} = 10.9$. 
When multiple ions relax simultaneously,
as in Figure \ref{fig:Goldman}(f), the charging dynamics are no longer
governed by a single parameter. Figure \ref{fig:Goldman} (h) shows that the system relaxes to the Goldman potential based on the combination of both timescales, with the fast dynamics determined by $\mbox{K}^{\tiny{+}}$ and the slower dynamics determined by $\mbox{Na}^{\tiny{+}}$. Hyper-polarization is observed as a direct consequence of the two relaxation timescales. The observed behavior is consistent with the refractory period in real cells. Defining the timescale to achieve stationary conditions as $t^*$, the final ion concentrations determined from simulation are listed in Table 1.

\section*{Summary}

We develop a mesoscopic representation for membrane diffusion based on lattice Boltzmann methods.
The method is designed to complement experimental imaging protocols,
so that non-equilibrium responses can be understood in the context of real cell geometries.
The method provides way to model how protein-mediated 
diffusion phenomena influence cellular responses based on a coarse-grained representation,
providing a way to model cellular systems at biologically relevant length and timescales.
Membrane transport coefficients are theoretically linked with the membrane diffusion coefficient for each chemical species, which can be measured or inferred from other experimental and computational approaches. Given a particular ion concentration field, chemical and electrical responses can be determined based on particular membrane transport properties. Transport in the bulk regions is determined from the Nernst-Planck Equations and Gauss's law using a coupled solution procedure. 
Inputs for the model are the initial concentration field and closure relationships for the membrane 
permeability. Transient responses for the concentration field and electrical potential
are outputs from the simulation.

Our approach is able to recover both equilibrium and non-equilibrium behavior
from first principles. We verify that simulations are able to recover the Nernst reversal potential when the membrane is permeable to a single ion, and the Goldman potential when the membrane is permeable to multiple ions. We show that the approach can be used to identify time constants associated with membrane charging, and that simulations for multi-ion transport can predict non-linear membrane dynamics. Hyper-polarization is observed based on the charging dynamics for multiple ions relaxing at different timescales. The formulation can support many different applications for lattice Boltzmann methods,
and consider wide range of biological and engineered systems. 

\section*{Author's Contributions}

\noindent Both authors contributed equally to this work. 

\begin{acknowledgments}
\noindent This research used resources of the Oak Ridge Leadership Computing Facility at the Oak Ridge National Laboratory, which is supported by the Office of Science of the U.S. Department of Energy under Contract No. DE-AC05-00OR22725. This research was also undertaken with the assistance of resources and services from the National Computational Infrastructure (NCI), which is supported by the Australian Government.
\end{acknowledgments}



\bibliography{References}

\end{document}